\documentstyle[12pt,eqsecnum,aps,epsfig]{revtex}

\begin{document}

\vbox{
\rightline{UTEXAS-HEP-98-14}

\vspace{1.0cm}

\title{First Observation of the Rare Decay Mode 
	$K^0_L \rightarrow e^+e^-$}

\author{
D. Ambrose$^{1}$,
C. Arroyo$^{2}$,
M. Bachman$^{3}$,
P. de Cecco$^{3}$,
D. Connor$^{3}$,
M. Eckhause$^{4}$,
K. M. Ecklund$^{2}$,
S. Graessle$^{1}$,
A. D. Hancock$^{4}$,
K. Hartman$^{2}$,
M. Hebert$^{2}$,
C. H. Hoff$^{4}$,
G. W. Hoffmann$^{1}$,
G. M. Irwin$^{2}$,
J. R. Kane$^{4}$,
N. Kanematsu$^{3}$,
Y. Kuang$^{4}$,
K. Lang$^{1}$,
R. Lee$^{3}$,
R. D. Martin$^{4}$,
J. McDonough$^{1}$, 
A. Milder$^{1}$,
W. R. Molzon$^{3}$,
M. Pommot-Maia$^{2}$,
P. J. Riley$^{1}$,
J. L. Ritchie$^{1}$,
P. D. Rubin$^{5}$,
V. I. Vassilakopoulos$^{1}$,
C. B. Ware$^{1}$,
R. E. Welsh$^{4}$,
S. G. Wojcicki$^{2}$,
E. Wolin$^{4}$,
S. Worm$^{1}$ \\
(BNL E871 Collaboration)\\
}

\address{ 
   $^{(1)}$University of Texas, Austin, Texas, 78712,\\
   $^{(2)}$Stanford University, Stanford, California, 94305,\\
   $^{(3)}$University of California, Irvine, California, 92697,\\ 
   $^{(4)}$College of William and Mary, Williamsburg, Virginia, 23187, \\
   $^{(5)}$University of Richmond, Richmond, Virginia, 23173\\ 
}

\date{\today}

\maketitle

\vspace{1.5cm}

\begin{abstract}

In an experiment designed to search for and study very
rare two-body decay modes of the $K^0_L$, we have observed
four examples of the decay $K^0_L \rightarrow
e^+e^-$, where the expected background is $0.17 \pm 0.10$ events.
This observation translates into a branching fraction of
$8.7^{+5.7}_{-4.1} \times 10^{-12}$,
consistent with recent
theoretical predictions.
This result represents by far the smallest
branching fraction yet measured in particle physics.

\end{abstract}

\vspace{6.0cm}

\noindent Submitted to Physical Review Letters.

}  

\newpage



We report the first observation of the
decay $K^0_L \rightarrow e^+ e^-$, which is strongly
suppressed by the GIM  mechanism~\cite{ref.n1} and the
helicity structure of the $V-A$ interaction. Thus, it
could be sensitive to a new physics process not subject to these
effects.
The rate of the analogous decay $K^0_L \rightarrow \mu^+ \mu^-$
is dominated by the absorptive contribution 
which arises from
the real two-photon intermediate state,
$K^0_L \rightarrow \gamma \gamma \rightarrow \mu^+ \mu^-$.
The absorptive contribution~\cite{ref.sehgal}
suggests a
$K^0_L \rightarrow e^+ e^-$
branching fraction  of
3$\times$10$^{-12}$.  However, recent
predictions~\cite{ref.n4} in the framework of
chiral perturbation theory
indicate that the two-photon
dispersive contribution is larger than the absorptive part
in the $K^0_L \rightarrow e^+ e^-$ decay.  The predicted branching 
fraction is approximately $9 \times 10^{-12}$.
The observation reported here is
consistent with these
predictions and consequently rules
out a significant 
non-Standard Model contribution to this decay.

Data were recorded in 1995 and 1996
by Experiment~871 at the Alternating Gradient
Synchrotron (AGS) at Brookhaven National Laboratory
(BNL).  The experiment built upon experience with
BNL~E791,
which set the best upper limit on 
$K^0_L \rightarrow e^+ e^-$~\cite{ref.e791}.
An improved search was made possible by the large increase
in AGS intensity with the new Booster 
and by apparatus modifications
to take advantage of higher flux while
improving background rejection.
Major parts of the spectrometer were 
more finely segmented to accommodate higher rates.  Other
important changes
were:
(a) redundancy was increased in the most critical areas
by making  three position measurements in all
tracking stations;
(b) magnetic fields were set to provide tracks nearly
parallel to the beam for 
two-body decays at the Jacobian peak; and
(c) a beam stop~\cite{ref.n6} was installed in the first
spectrometer magnet to absorb the neutral beam.
Taken together, these modifications made possible an improvement
of roughly an order of magnitude over the earlier experiment.

\begin{figure}
 \begin{center}
 \epsfig{ file=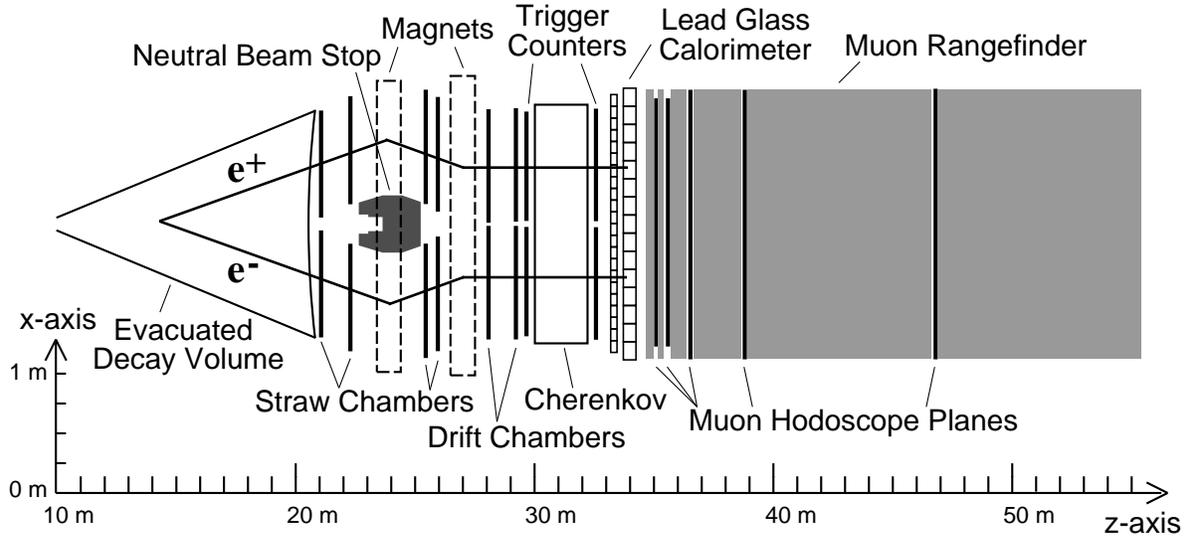, width=6.3in} 
 \end{center}
 \caption{Plan view of the E871 apparatus.
The origin of the $z$-axis 
is at the target.}
 \label{v6fig_1}
 \end{figure}

The neutral beam was produced by 24~GeV protons
on a 1.4 interaction length Pt target at
$3.75^\circ$ with respect to the collimation channel.  Typical 
proton intensity was $1.5 \times 10^{13}$
per slow spill (1.2--1.6~s
duration).  This provided about
$2 \times 10^8$ $K^0_L$ per spill ($2 < p_K < 16~{\rm GeV/c}$)
at the entrance of the decay volume, of which 
7.5\% decayed in the 11~m fiducial length.
The experiment is shown in Fig.~\ref{v6fig_1}.
For brevity, only the
detectors most relevant to this analysis will
be discussed.  These are tracking, electron
identification, and trigger hodoscopes.

The tracking section consisted of six chamber
stations and two consecutive dipole magnets.
The magnets had opposite polarities and provided
net transverse momentum kicks of 418~MeV/c and 216~MeV/c.  
The topology of a $K^0_L \rightarrow e^+ e^-$ decay
is illustrated in Fig.~\ref{v6fig_1}.
The upstream four tracking stations,
where the highest rates occurred, 
were constructed from 5~mm diameter
straw tubes~\cite{ref.n7}
and were operated with a fast gas mixture (50-50 CF$_4$/C$_2$H$_6$).
The downstream-most
two stations were in a region of lower
rates and consisted of
drift chambers with 1~cm sense wire spacing operated 
with a 50-50~Ar/C$_2$H$_6$ gas mixture.  
Each tracking station provided three measurements of position
in the horizontal (magnet-bend) $x$-plane and two measurements
in the vertical $y$-plane.
Since the resolution of the system
was limited mainly by multiple Coulomb scattering,  
the total material 
was kept to a minimum;
including helium bags between the
chambers, it amounted to
$1.5 \times 10^{-2}$ radiation lengths. 
The
single wire resolutions were 160 and 120~$\mu$m,
and 
the average efficiencies (per wire)
were
96\% and
98\%, for the straw and drift chambers, respectively. 
The measured mass resolutions for 
$K^0_L \rightarrow \pi^+ \pi^-$
and $K^0_L \rightarrow \mu^+ \mu^-$ 
decay modes were 1.11 and 1.28~MeV/c$^2$, 
respectively, compared with the 1.07
and 1.22~MeV/c$^2$ predicted by Monte
Carlo.  The $K^0_L \rightarrow e^+ e^-$
mass resolution 
predicted by Monte Carlo is 1.39~MeV/c$^2$.

Redundant electron identification was achieved by an
atmospheric hydrogen
threshold
Cherenkov  counter (CER) and a lead glass array (PBG).
The CER had 4-by-4 arrays of mirror-phototube pairs on each
side of the beam. 
The average
photoelectron yield was 5.5 for electrons. 
The PBG consisted of 216 blocks arranged in  two
layers, with 3.5~radiation length converter 
blocks in front of  10.5~radiation length absorber blocks.
The measured PBG energy resolution for electrons was
$\sigma/E = 0.015 + 0.062/\sqrt{E(\rm GeV)}$.  
The CER and PBG performance
is summarized in Table~\ref{table_1}.

Two scintillator hodoscopes, one upstream 
of the CER and one downstream (separated by
2.9~m), were used for triggering.
Both hodoscopes had 3.2~cm wide  $x$-measuring slats 
with
phototubes on both ends.  The downstream hodoscope had, in
addition, 3.0~cm wide $y$-measuring slats with a phototube on
one end.
The slats 
overlapped their nearest neighbors by 3~mm to avoid
inefficiency due to cracks.

Online data selection involved hardware 
and software triggers.
A Level~0 (L0)
trigger based on the pattern of trigger hodoscope hits
required two in-time tracks to be  parallel
within about 30~mrad to, and on opposite sides of, the
beam axis.
The L0 trigger along with signals from particle identification
detectors formed  Level~1 (L1) triggers for various
di-lepton modes.
The L1 trigger for $e^+e^-$   
required CER hits to be
in time and spatially correlated with the trigger hodoscope hits.
PBG signals were not required.
For events passing a L1 trigger,
a software algorithm, 
run on a farm of eight RISC processors, constituted a
Level~3 (L3) trigger.  The algorithm
reconstructed tracks using information from the 
trigger hodoscopes and 
all tracking chambers.
For $e^+ e^-$ events, it required tracks on
each side of the spectrometer which formed a vertex in the
decay volume.
Also,  L0 triggers prescaled 
by a 
factor of 1000 were
recorded for analysis.  This minimum bias sample
was used to study detector performance and provided
data for normalization.

The spectrometer had a geometrical acceptance of 1.89\% for
$K_L^0 \rightarrow e^+ e^-$ decays with $9.75 < z < 20.75~{\rm m}$
(see Fig.~1) and kaon momenta between 2 and 16~GeV/c.
The trigger requirement of parallelism reduced the acceptance to
1.57\%.  Analysis criteria
further reduced the acceptance to 1.23\%.

Potential sources of 
background 
include 
misidentified  
$K^0_L \rightarrow \pi^\pm e^\mp\nu$ decays, 
accidental spatial and temporal coincidences
of $e^+$ and $e^-$ from two  $K^0_L \rightarrow \pi^\pm e^\mp\nu$ decays,
partially measured 
 $K^0_L \rightarrow e^+ e^- \gamma$,
$K^0_L \rightarrow e^+e^-e^+e^-$, or
 $K^0_L \rightarrow e^+ e^- \gamma \gamma$ decays,
and  $K^0_L \rightarrow \gamma \gamma$ decays 
with asymmetric external
conversion of both photons in the vacuum window or first straw
chamber.
The only sources which were not negligible
after all analysis criteria were applied
are  $K^0_L \rightarrow e^+e^-\gamma$\cite{ref.eeg} and
$K^0_L \rightarrow e^+e^-e^+e^-$~\cite{ref.eeee} decays.
Both of these decays have a low
probability of
producing an $e^+e^-$ pair with invariant mass
near the $K_L^0$ mass.

Analysis procedures were designed to ensure
that selection criteria 
were not influenced by knowledge of events in or
near the signal region.
$K^0_L \rightarrow e^+e^-$ events should have measured $e^+ e^-$
mass near the
$K^0_L$ mass (497.7~MeV/c$^2$)
and  measured transverse momentum ($p_T$), defined with respect to
the parent $K^0_L$ direction of flight,
near zero.
Thus, events with $490 < M_{ee} < 505$~MeV/c$^2$ 
and $p_T^2 < 100$~(MeV/c)$^2$ were excluded from consideration
until all selection criteria were finalized.

As a first step in the analysis, events were  required 
to have a
good track 
(with signals from at least two  $x$-measuring wires and
one $y$-measuring wire in each chamber)  
on
each side of the spectrometer.  These tracks had to form
a good vertex in the decay volume and project to in-time
trigger counter hits consistent with the
parallelism requirement.  
Two independent fitting
algorithms, which used a full magnetic field map, 
subsequently determined
kinematic quantities for each track.  The fitters had different
sensitivities to track finding errors and 
checked each other's results.  Ultimately,
all selection criteria were applied
to the results of one fitter, with the exception that consistent
results for $M_{ee}$ and $p_T^2$ were required of both.

Events had to
satisfy minimal track and vertex quality
criteria, 
be fully contained within the spectrometer,
and have $M_{ee} > 475$~ MeV/c$^2$ 
and $p_T^2 < 900$~(MeV/c)$^2$ 
(excluding the region 
$490 < M_{ee} < 505$~MeV/c$^2$ 
and $p_T^2 < 100$~(MeV/c)$^2$).
In addition, events had to have energy deposition in
the PBG characteristic of electrons
and good CER pulse height and timing information.  
The remaining 833 events 
were studied to determine criteria to further
exclude background with minimal acceptance loss.
This optimization was based on Monte Carlo and studies of
observed
$K^0_L \rightarrow \pi^+ \pi^-$ events.
These criteria included better
track and vertex quality, tighter
timing,  a requirement that the momentum
asymmetry between the tracks 
($|p_{e^+} - p_{e^-}|/(p_{e^+}  + p_{e^-})$)
be less than 0.55 (to suppress  
$K^0_L \rightarrow \pi^\pm e^\mp\nu$ decays),
and a requirement that no additional complete tracks 
be found in the spectrometer (to suppress events with
two $K_L^0$ decays).  
Finally, to reduce $K^0_L \rightarrow e^+ e^- e^+ e^-$ background,
criteria were applied to reject 
events 
with short partial tracks in the
two upstream-most straw chambers pointing to
the reconstructed $e^+ e^-$ vertex.
The resulting sample comprised 44 events.

After all
analysis criteria were finalized, a signal region was defined
(elliptical in $p_T^2$ and $M_{ee}$ 
and corresponding to about 2.5~sigma in each).  
As a result of
inner bremsstrahlung~\cite{ref.ib},
23\% of $K_L^0 \rightarrow e^+ e^-$ decays fall outside this
ellipse.
The size of the signal region was chosen to reduce the expected level
of background to well below one event.
Background estimates were based on Monte Carlo simulations and
comparisons to the data.
Samples of $K^0_L \rightarrow e^+ e^- e^+ e^-$
and $K^0_L \rightarrow e^+ e^- \gamma$
decays were generated~\cite{ref.gen}.
These distributions were absolutely normalized 
by their measured branching fractions~\cite{ref.pdg}
to obtain  predictions of 
$38 \pm 8$ events from $K^0_L \rightarrow e^+ e^- \gamma$
and 
$24 \pm 11$ events from $K^0_L \rightarrow e^+ e^- e^+ e^-$
in the region defined by $ 476 < M_{ee} < 490~{\rm MeV/c}^2$
and $p_T^2 < 400$ (MeV/c)$^2$,
where  43~events were observed.  
The uncertainties in the 
predictions are mainly
due to the $K^0_L \rightarrow e^+ e^- \gamma$ form factor and
the fact that the 
$K^0_L \rightarrow e^+ e^- e^+ e^-$
efficiency of
the partial track cut is not well known.
To remove these sources of systematic 
uncertainty from the estimate of
the background in the signal region,
we performed a fit to data to normalize the background
distributions, which were then extrapolated into the signal region.
The fit in $p_T^2$ and $M_{ee}$ exploited  the fact that
in the range $ 476 < M_{ee} < 490~{\rm MeV/c}^2$
the $p_T^2$ distributions
for the background decays differ significantly.  
This procedure yielded estimates of background in the signal region
of $0.09 \pm 0.11$~event from $K^0_L \rightarrow
e^+e^-e^+e^-$, $0.08 \pm 0.02$~event from 
$K^0_L \rightarrow e^+ e^- \gamma$, and
$0.17 \pm 0.10$ for the sum of both
(to be compared with
$0.37 \pm 0.14$
for the absolutely normalized prediction).
  
\begin{figure}
 \begin{center}
 \epsfig{ file=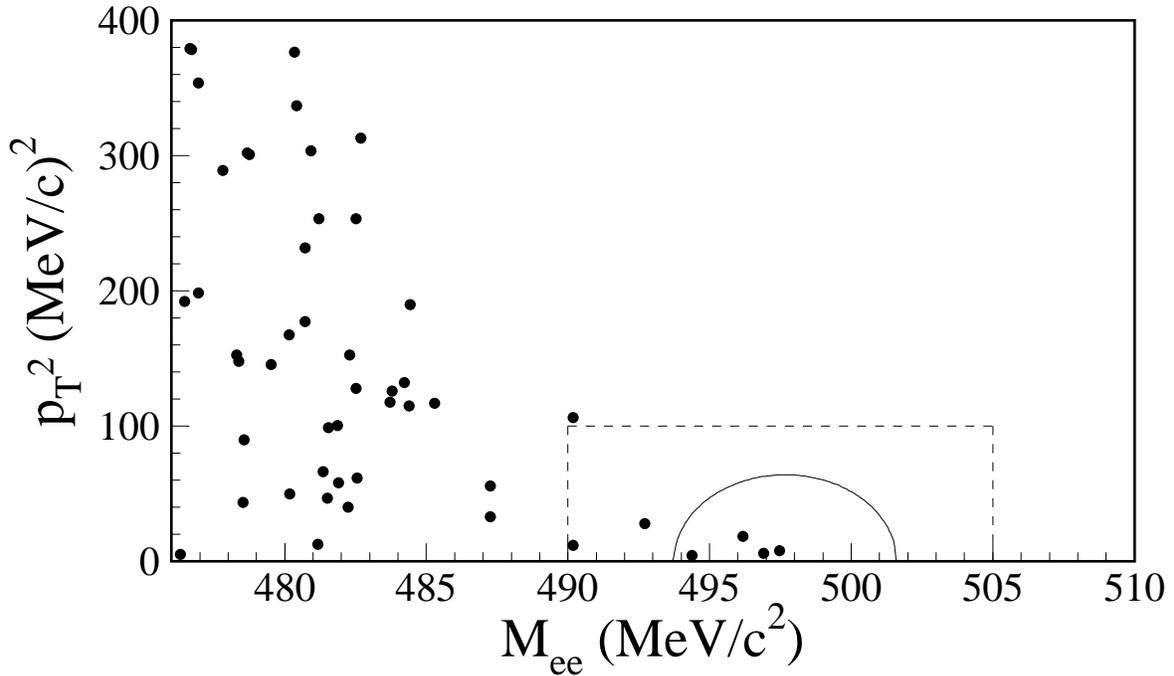, width=6.3in }
 \end{center}
 \caption{$p_T^2$ versus $M_{ee}$ for $K^0_L \rightarrow e^+ e^-$
candidates.  The dashed line shows the exclusion region.
The solid curve bounds the signal region.
}
 \label{v6fig_2}
 \end{figure}

When the full analysis was performed on the excluded region,
four $K_L^0 \rightarrow e^+ e^-$
candidates were found inside the
signal ellipse, as shown in Fig.~\ref{v6fig_2}.
These candidates have been carefully scrutinized and
exhibit no anomalous features.  
The probability of observing 4~background events in the signal region
when 0.2 are expected
is $6 \times 10^{-5}$.
A check on the reconstructed mass of
$K^0_L \rightarrow \pi^+ \pi^-$  events collected
within minutes of
the $K^0_L \rightarrow e^+ e^-$ candidates 
rules out transient shifts in the mass scale.

\begin{figure}
\begin{center}
\epsfig{ file=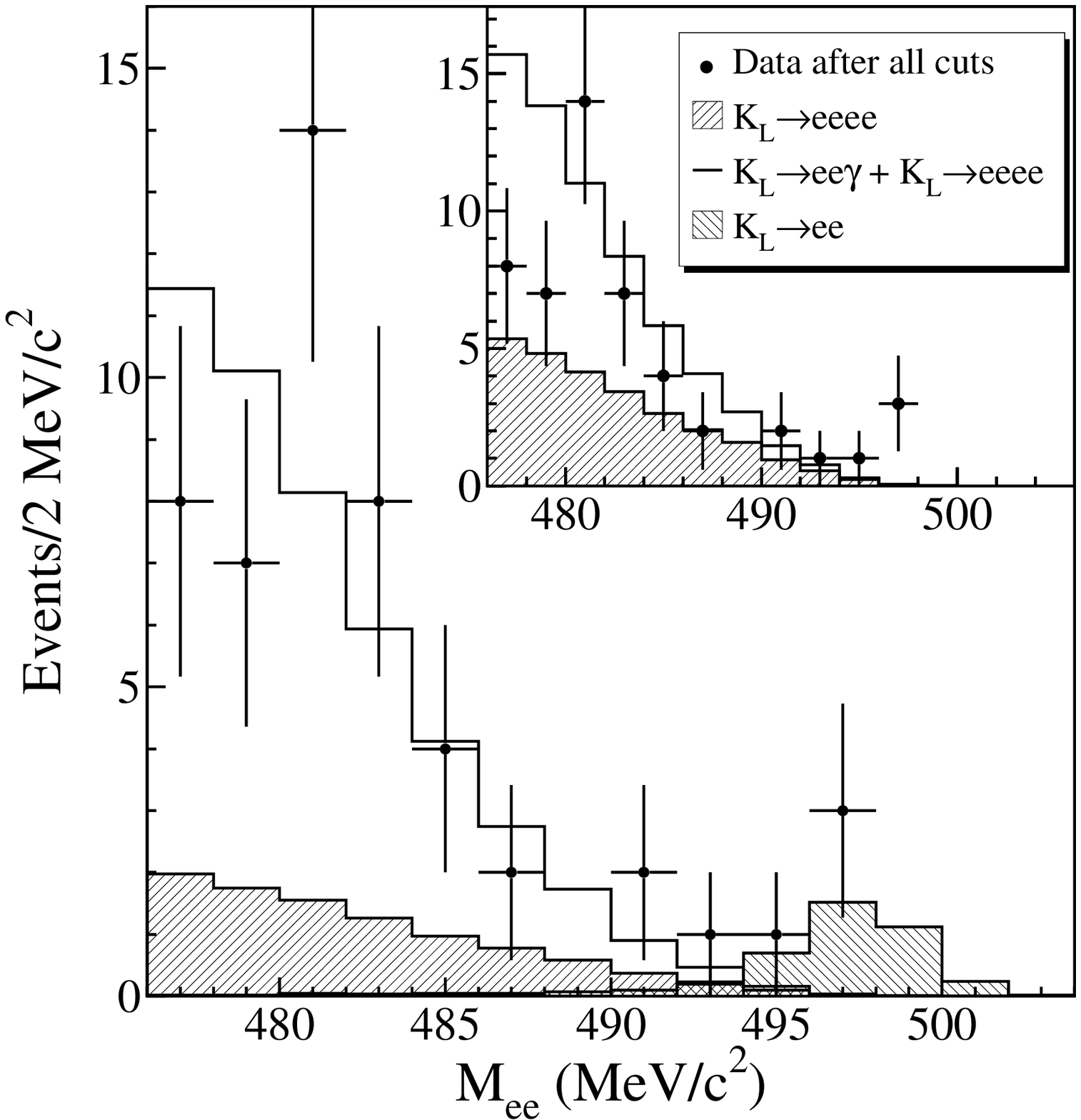, width=6.3in }
\end{center}
\caption{
Results of the maximum likelihood fit, showing the distributions of
$K^0_L \rightarrow e^+ e^-$,
$K^0_L \rightarrow e^+e^-e^+e^-$,
and $K^0_L \rightarrow e^+ e^- \gamma$  
along with the data versus $M_{ee}$.
The inset shows the absolute Monte Carlo prediction for the 
background distributions, which have systematic uncertainties described
in the text.
}
\label{v6fig_3}
\end{figure}

To extract the best estimate of the 
number of $K_L^0 \rightarrow e^+ e^-$ events, a maximum
likelihood fit was performed
in the region defined by $476 < M_{ee} < 510$ MeV/c$^2$
and $p_T^2 < 400$ (MeV/c)$^2$.  
In addition to the 50 events
inside this region, the inputs to the fit were
Monte Carlo distributions in $M_{ee}$ and $p_T^2$ for
$K^0_L \rightarrow e^+e^-e^+e^-$,
$K^0_L \rightarrow e^+ e^- \gamma$,
and $K^0_L \rightarrow e^+ e^-$ decays. 
The fit estimated the number of events in each distribution,
subject to the constraint that the sum equal the observed number of
events.
Fig.~\ref{v6fig_3} shows the fit results for the three
distributions versus $M_{ee}$.
The result for the number of $K^0_L \rightarrow e^+ e^-$
events is $4.20^{+2.69}_{-1.94}$ in the full region of the fit.
For comparison, the inset in Fig.~\ref{v6fig_3} shows the
$M_{ee}$ distribution of data along with the 
absolutely normalized Monte Carlo predictions for
$K^0_L \rightarrow e^+e^-e^+e^-$ and
$K^0_L \rightarrow e^+ e^- \gamma$.

The $K^0_L \rightarrow e^+e^-$  branching fraction $B_{ee}$  
was
determined 
from the formula

\[
B_{ee} = N_{ee}\,
	\frac{ B_{\pi \pi}}
	{R  N_{\pi\pi}} \,
	\frac{A_{\pi\pi}}{A_{ee}} \,
	\frac{1}{\epsilon^{L1}_{ee}} \,
	\frac{1}{\epsilon^{L3}_{ee}} \,
        \frac{1}{\epsilon^{PID}_{ee}} f_{\pi \pi}.
\]

\noindent
$N_{ee}$ is the number of $K^0_L \rightarrow e^+ e^-$ 
events determined from
the likelihood fit.
$B_{\pi \pi}$ is the 
$K^0_L \rightarrow \pi^+ \pi^-$ branching fraction~\cite{ref.pdg}.
$R$ is the prescale factor for the minimum
bias sample used for normalization.  It was the product of the
hardware prescale (1000) and an additional factor 
of 20 imposed in software.
$N_{\pi\pi}$ is the number of
$K^0_L \rightarrow \pi^+ \pi^-$ events 
in this reduced sample  (and includes a
small correction for the effect of $K_S^0$ interference).  
The ratio $A_{\pi\pi}/A_{ee}$, calculated via Monte Carlo,
corrects for mode-dependent acceptance differences 
due to the 
detector geometry, 
trigger, 
and analysis criteria.
$\epsilon^{L1}_{ee}$ is the L1 $ee$ trigger efficiency with respect
to the L0 trigger (which all events had to satisfy).
$\epsilon^{L3}_{ee}$ is the L3 $ee$ trigger  efficiency
and was derived from the measured L3 efficiency for
$K^0_L \rightarrow \pi^+ \pi^-$ by making a small correction
to account for the different kinematics of $K_L^0 \rightarrow e^+ e^-$.
$\epsilon^{PID}_{ee}$ is the particle identification efficiency
and was determined by weighting measured detector efficiencies
by the appropriate distributions for 
$K^0_L \rightarrow e^+ e^-$ decays. 
$f_{\pi \pi}$ accounts for the loss of
$K^0_L \rightarrow \pi^+ \pi^-$ events due
to 
hadronic interactions in the spectrometer and was 
based on a GEANT calculation.
The calculation was checked against
special data taken with a single-arm trigger, which provided 
events in which
a pion was lost due to hadronic interaction in the non-trigger arm.
The values of these
parameters 
are given in Table~\ref{table_2}. 
The
calculated branching fraction $B_{ee}$ is 
$8.7^{+5.7}_{-4.1} \times 10^{-12}$.

In summary, we have made the first observation
of the decay $K^0_L
\rightarrow e^+e^-$.
The measured branching fraction is consistent
with recent theoretical predictions~\cite{ref.n4}
and is the
smallest measured 
in particle physics to date.


   We acknowledge the support of the BNL staff, particularly
H.~Brown, R.~Brown, R.~Callister, A.~Esper,
F.~Kobasiuk, W.~Leonhardt, M.~Howard, J.~Negrin,
and J.~Scaduto. 
M.~Hamela and
D.~Ouimette were key in the development of straw
chambers and electronics, as were 
S.~Kettell and R.~Atmur in the development of the
L1 and L3 triggers, respectively.
C.~Allen, G.~Bowden,
P.~Coffey, M.~Diwan, M.~Marcin, C.~Nguyen, and
A.~Schwartz made important contributions in early design and
construction of the experiment. 
We thank V.~Abadjev,  
P.~Gill, N.~Mar,
J.~Meo, 
M.~Roehrig, and M.~Witkowski 
for valuable technical assistance. 
We thank the SLAC Computing Division and the
BNL CCD for help with data processing. 
Finally, we thank D.~Dicus for help with radiative
corrections.
This work was supported in
part by the U.S. Department of Energy, the National Science Foundation,
the Robert A. Welch Foundation, and Research Corporation.

 \clearpage

 \begin{table}
 \caption{ Performance of particle identification 
detectors as measured from a clearly
   identified sample of $K_{e3}$ and $K_{\mu3}$ events. 
The Cherenkov $\pi$ and $\mu$ 
rejections are
calculated for particles with momenta
below their Cherenkov thresholds,
8.3~GeV/c and 6.3~GeV/c, respectively.}
 \label{table_1}
 \begin{tabular}{lcc}
     & Cherenkov  & Lead Glass  \\
 \tableline
 $e$ efficiency  & $0.977 \pm$ 0.001 & $0.987 \pm 0.004$ \\
 $\pi$ misidentification & $0.0019 \pm 0.0002  $ 
                         & $0.0093 \pm 0.0004$ \\
 $\mu$ misidentification & $0.0024 \pm 0.0002 $ 
                         & $0.0018 \pm 0.0002$ \\
 \end{tabular}
 \end{table}

 \begin{table}
 \caption{ Factors entering into the calculation of the 
$K^0_L \rightarrow e^+ e^-$ branching fraction.}
 \label{table_2}
 \begin{tabular}{ c c c c c }
\qquad & Variable & \qquad & Value & \qquad \\
 \tableline
& $N_{ee}$  		&  & $4.20_{-1.94}^{+2.69}$ &  \\
& $B_{\pi\pi}$~\cite{ref.pdg} 
			& & $ (2.067 \pm 0.035) \times 10^{-3}$ & \\
& $R$                    & & $ 20000            $   &      \\
& $N_{\pi\pi}$		& & $ 83531  \pm 381   $    &     \\
& $A_{\pi\pi}/A_{ee}$	& & $ 1.478  \pm 0.011 $    &     \\
& $\epsilon^{L1}_{ee}$	& & $ 0.977  \pm 0.011 $  &	\\
& $\epsilon^{L3}_{ee}$	& & $ 0.929  \pm 0.003 $  & 	\\
& $\epsilon^{PID}_{ee}$  & &  $ 0.929  \pm 0.011 $ & 	\\
& $f_{\pi\pi}$           & & $ 0.954  \pm 0.004 $   &      \\
 \end{tabular}
 \end{table}

  \psfigdriver{dvips}

\begin{references}


\bibitem{ref.n1} S.L. Glashow, J. Iliopoulos, and L. Maiani, 
Phys.\ Rev.\ D. {\bf 2},  1285 (1970).
\bibitem{ref.sehgal} L.M. Sehgal, Phys.\ Rev.\ {\bf 183},
                1511 (1969).
\bibitem{ref.n4} G. Valencia, Nucl.\ Phys.\ {\bf B517}, 339 (1998);
 D. Gomez Dumm and A. Pich, Phys.\ Rev.\ Lett.\ {\bf 80}, 4633  (1998).
\bibitem{ref.e791} K. Arisaka {\em et al.}, Phys.\ Rev.\ Lett.\ 
{\bf 71}, 3910 (1993).
\bibitem{ref.n6} J. Belz {\it et al.}, {\it A Compact Beam Stop for
a Rare Kaon Decay Experiment,} hep-ex/9808037,
submitted to Nucl.\ 
Instrum.\ Methods.
\bibitem{ref.n7} S. Graessle {\em et al.},
Nucl.\ Instrum.\ Methods, 
{\bf A 137}, 138 (1995).

\bibitem{ref.eeg} 		
G.D. Barr {\em et al.}, Phys. Lett. {\bf B 240}, 283 (1990);
K.E. Ohl {\em et al.}, Phys. Rev. Lett. {\bf 65}, 1407 (1990);
W. M. Morse {\em et al.}, Phys. Rev. {\bf D 45}, 36 (1992).

\bibitem{ref.eeee}
T. Akagi {\em et al.}, Phys. Rev. {\bf D 47}, 2644 (1993);
M.R. Vagins {\em et al.}, Phys. Rev. Lett. {\bf 71}, 35 (1993);
P. Gu {\em et al.}, Phys. Rev. Lett. {\bf 72}, 3000 (1994);
G.D. Barr {\em et al.}, Z. Phys. {\bf C 65}, 361 (1995).

\bibitem{ref.ib} Inner bremsstrahlung 
is included in the Monte Carlo
using the results of D.A. Dicus and W.W. Repko (to be published), 
which  are
consistent with
L. Bergstrom, Z. Phys., {\bf C20}, 135 (1983) for the $e^+e^-$ mode.

\bibitem{ref.gen} 
Generation of $K_L^0 \rightarrow e^+ e^- \gamma$ was based on 
N. Kroll and W. Wada, Phys.\ Rev. {\bf 98}, 1355 (1955),
modified by a form factor from 
L. Bergstrom, E. Masso, 
and P. Singer, Phys.\ Lett. {\bf 131B}, 229 (1983),
with value $\alpha_{K^*} = -0.28$ (as given by 
Reference~\cite{ref.pdg}).  
Generation of  $K^0_L \rightarrow e^+e^-e^+e^-$
was based on  T. Miyazaki and E. Takasugi, Phys.\ Rev.\  {\bf D 8},
2051 (1973). 
The introduction of a form factor (according to
L. Zhang and J.L. Goity, Phys.\ Rev. {\bf D57}, 7031 (1998)) 
had negligible effect on our
analysis.

\bibitem{ref.pdg} Particle Data Group, Eur.\ Phys.\ J. {\bf C 3}, 1 (1998).


\end{references}
\end{document}